\documentstyle[prl,aps,multicol,epsfig]{revtex}

\begin{document}
\draft

\title{Dielectric catastrophe at the Mott transition}

\author{C.\ Aebischer, D.\ Baeriswyl, and R.\ M.\ Noack}
\address{Institut de Physique Th\'eorique, Universit\'e de Fribourg,
      	 CH--1700 Fribourg, Switzerland}
\maketitle

\begin{abstract}
We study the Mott transition as a function of interaction strength in
the half--filled Hubbard chain with next--nearest--neighbor hopping
$t'$ by calculating the response to an external electric field using
the Density Matrix Renormalization Group.
The electric susceptibility $\chi$ diverges when approaching the
critical point from the insulating side. 
We show that the correlation length $\xi$ characterizing this
transition is directly proportional to fluctuations of the polarization 
and that $\chi\sim\xi^2$.
The critical behavior shows that the transition is
infinite--order for all $t'$, whether or not a spin gap is present, 
and that hyperscaling holds.
\end{abstract}

\pacs{PACS numbers: 71.10.Fd, 71.30.+h, 75.40.Cx}

\begin{multicols}{2}
\narrowtext

A material's response to an applied electric field characterizes
whether it is a metal or an insulator.
One such response is the static electrical conductivity at zero
temperature, which is finite for a metal (or infinite for an ideal
conductor), but vanishes for an insulator \cite{kohn}.
The conductivity can therefore be used to probe the metal--insulator
transition from the metallic side.
A complementary quantity is the dielectric response to an electric field,
the electric susceptibility, $\chi$. 
This quantity is expected to diverge (for a continuous transition) 
when the transition is approached
from the insulating side and to remain
infinite in the metallic phase. 
This phenomenon, termed ``dielectric catastrophe'' by Mott
\cite{mott}, has been reported for doped silicon \cite{rosenbaum}.

One possible origin of insulating behavior is the local Coulomb
repulsion  between electrons.
This ``Mott phenomenon'' \cite{anderson} leads to a
metal--insulator transition which occurs either as the
electron density, $n$, is varied for fixed electron--electron interaction
strength or as a function of interaction strength at fixed electron
density\cite{mott,gebhard}.
In this letter, we concentrate on the transition as a function of
interaction strength for fixed electron density.
Experimentally, such a transition can be induced by applying isostatic
or chemical pressure.

The prototype model for the Mott transition is the
single--band Hubbard model with purely local interaction, whose
Hamiltonian is
\begin{equation}
  \hat{H}=-\sum_{i j \sigma} t_{ij}\;\hat{c}_{i\sigma}^\dagger 
  \hat{c}_{j\sigma}
   +U\sum_i \hat{n}_{i\uparrow}\hat{n}_{i\downarrow} \;	, 
\label{eqn:Hubbard}
\end{equation}
where $\hat{c}_{i\sigma}^\dagger$ creates an electron of spin $\sigma$
at site $i$ and 
$\hat{n}_{i\sigma} \equiv \hat{c}_{i\sigma}^\dagger \hat{c}_{i\sigma}$.
The hopping matrix elements $t_{ij}$ are short--ranged. 
At half--filling, $n=1$, the Hamiltonian (\ref{eqn:Hubbard}) maps onto
a Heisenberg model with couplings $J_{ij}=4t_{ij}^2/U$
for $U\rightarrow\infty$ and is thus insulating, while at $U=0$,
it describes a perfect metal. 
Therefore, a Mott transition must occur at some $U_c\geq0$
\cite{sachdev:book}.
 
In order to describe the dielectric response of such a system,
one must consider the coupling to a static electric field.  
Taking the field in the $x$--direction
and neglecting overlaps between different Wannier functions
(tight--binding approximation), we add
the coupling term
\begin{equation}
\hat{H}_{ext} = -E\hat{X} = -E \sum_i x_i \hat{n}_i\ ,
\label{coupling}
\end{equation}
where $\hat{X}$ is the dipole operator (we have put $q=1$), $x_i$ is
the $x$--coordinate of the $i$-th site and $\hat{n}_i$ measures the
occupation of this site.
Here we have assumed that the finite lattice has 
{\it open boundary conditions}, i.e. the connections terminate at the
lattice edges.
We note that while this is the natural definition for experiments,
the notion of response to an applied electric field has recently been
generalized to periodic boundary conditions \cite{resta}.
An applied electric field will induce a polarization
at zero temperature given by
\begin{equation}
P=L^{-d} \langle X\rangle \ = \ -L^{-d} \ \frac{\partial E_0}{\partial E}
\label{polarization}
\end{equation}
on a $d$--dimensional lattice with linear dimension $L$, 
where the average is taken with respect to the ground state of the
full Hamiltonian $\hat{H} + \hat{H}_{ext}$\ , with corresponding
energy $E_0$. 
The zero--field susceptibility is then defined as
\begin{equation}
\chi\ = \ \left.{\frac{\partial P}{\partial E}}\right|_{E=0}\ = \
-L^{-d} \left.{\frac{\partial^2 E_0}{\partial E^2}}\right|_{E=0}\ \; .
\label{eqn:susceptibility}
\end{equation}
The examination of the properties of this susceptibility in the
vicinity of the Mott metal--insulator transition is the principle aim
of this letter.

The susceptibility $\chi$ can be
related to the eigenstates  $|\Psi_n\rangle$ of $\hat H$ using
elementary perturbation theory,
\begin{equation}
\chi\ = \ 2L^{-d}\sum_{n\not=0}\frac{|\langle\Psi_0|\hat X|\Psi_n\rangle|^2}
{\Delta E_n}\ ,
\label{eqn:perturbation}
\end{equation}
where $\Delta E_n$ is the excitation energy of the $n$-th eigenstate.
(Here we have chosen the origin of the coordinate system so that
$\langle X\rangle$ = 0 for $E=0$.)
This relation immediately yields a useful inequality
in terms of 
the ``charge gap'', $\Delta$ (defined as the lowest excitation energy
for which the dipole matrix element does not vanish):
\begin{equation}
\chi \ \le \ \frac{2}{\Delta}\ L^{-d}\langle\Psi_0|\hat X^2|\Psi_0\rangle\ .
\label{inequality}
\end{equation}
It is thus instructive to consider fluctuations of the polarization,
$\langle\Psi_0|\hat X^2|\Psi_0\rangle$,
which can be estimated as follows. 
We expand the ground state as a series
$|\Psi_0\rangle=\sum_D|\Psi_0^{(D)}\rangle$,  where $D$ is the
number of doubly occupied sites (``particles''). 
At large $U$ the ``particles'' are located close to empty sites
(``holes''). 
Each particle--hole pair represents an elementary dipole with
essentially random orientations.
Therefore our estimate is
\begin{equation}
\langle\Psi_0|\hat X^2|\Psi_0\rangle\ = \
\sum_D \langle\Psi_0^{(D)}|\hat X^2|\Psi_0^{(D)}\rangle
\ \approx \ \langle D \rangle \; l^2\ ,
\label{estimate}
\end{equation}
where $l$ is the average size of the dipoles. Comparing this result
with the inequality in Eq.\ (\ref{inequality}), we conclude that a
diverging susceptibility requires either a diverging size of the
dipoles or a vanishing charge gap or both. In one dimension,
the quantity
\begin{equation}
	  \xi=\frac{1}{L}\langle\Psi_0|\hat X^2|\Psi_0\rangle
	  \label{xi:def}
\end{equation}	
is a length characterizing the insulating phase 
\cite{kudinov:1,souza:wilkins}.
We will show below that $\xi$ is the {\em correlation length}, up to a
dimensionless constant.

On regular lattices, one often faces the problem that the Mott
phenomenon, which sets in at large values of $U$ due to charge
blocking, is masked by the opening of a charge gap at much lower
values of $U$ due to antiferromagnetic order induced by nesting
or Umklapp processes.
In order to control such effects, we consider here a model
that explicitly incorporates frustration of antiferromagnetism, namely
the one--dimensional Hubbard model with both nearest--neighbor 
$t$ and next--nearest--neighbor $t'$ hopping terms. We set $t=1$ and
consider only $t'\ge 0$ here because the sign of $t'$ is
irrelevant at half--filling due to particle--hole symmetry.
For $t'=0$, 
the Bethe--Ansatz solution allows one to calculate the
charge gap \cite{ovchin}, the charge stiffness, and the correlation length
in the insulator \cite{stafford} explicitly.
The system is found to be insulating for all positive values of $U$.
The metal--insulator transition occurs at $U_c=0$ and is infinite
order: the charge gap and, correspondingly, the inverse of the
correlation length decrease exponentially as $U\rightarrow 0^+$.
At the same time, the magnetic correlations show a power--law decay. 
For $t'>0$, a weak--coupling renormalization group analysis
\cite{fabrizio} predicts the same behavior as long as there are two
Fermi points: Um\-klapp processes lead to an insulating state for $U>0$,
while the magnetic excitation spectrum remains gapless.

For $t' > 0.5$, there are four Fermi points in the
noninteracting band structure and the picture becomes more complicated.
In weak coupling, the lowest--order Um\-klapp processes are marginally
irrelevant \cite{fabrizio}, and the system is predicted to be
metallic (vanishing charge gap) with a spin gap.
At strong coupling, the model can be mapped to a frustrated Heisenberg
chain, which develops a spin gap for  
$J'/J \sim t'^2 > 0.2412$ \cite{eggert} and incommensurate
antiferromagnetic order for $J'/J > 0.5$.
This general picture has been confirmed numerically\cite{kuroki,daul}.
For a detailed phase diagram, 
we refer the reader to Fig.\ 3 of Ref.\ \onlinecite{daul}.
Here we will examine both the parameter regime with gapless magnetic
excitations and $U_c = 0$ ($t' \lesssim 0.5$) and the one with gapped spin
degrees of freedom and $U_c > 0$ ($t' \gtrsim 0.5$).

In order to numerically evaluate the electric susceptibility, 
Eq.\ (\ref{eqn:susceptibility}), we use the Density Matrix
Renormalization Group (DMRG)\cite{white}.
We apply a small electric field so that the 
system is in a linear response regime (typically $EL = 0.001$)
and measure
\begin{equation}
  \chi=\frac{P}{E}=
         \frac{1}{L E} \sum_i x_i \; \langle \hat{n}_i\rangle \; .
\end{equation}
We use the finite--size DMRG algorithm \cite{white,noackwhite} on up
to $L=100$ sites, retaining up to 2400 states for the system block.
This allows us to keep the sum
of the discarded density--matrix eigenvalues to below $10^{-8}$.
We have performed extensive tests for $U=0$, a difficult case to treat
numerically, and find that we can reproduce
analytic results to within less than one percent.
The details of the calculations will be described more extensively
elsewhere.
\begin{figure}
\begin{center}
\epsfig{file=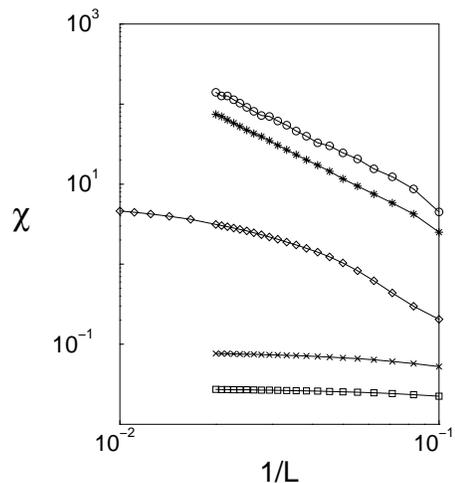,width=6cm}
\end{center}
\caption{
Electric susceptibility, $\chi$, as a function of $1/L$ for
$U=1$ (circles), $U=2.5$ (stars),  
$U=4$ (diamonds), $U=5.5$ (crosses), and $U=7$ (squares).
}
\label{fig:chi_L}
\end{figure}
The electric susceptibility $\chi$ is
shown in Fig.\ \ref{fig:chi_L} as a function of the inverse system
size for $t'=0.7$ and a number of $U$--values.
There are two characteristically dif\-ferent behaviors: at small $U$,
the system is metallic, and the susceptibility diverges with system
size.
A fit to a power law in $L$ yields an exponent very close to 2 (within
$5\%$) for
the small $U$ values.
For $U=0$, it can be shown analytically that $\chi \sim L^2$ for large
$L$ for all values of $t'$.
We conjecture that such a $L^2$ divergence of $\chi$ is {\em generic} for a
one--dimensional perfect metal.
For larger $U$, $\chi$ extrapolates to a finite value as 
$L \rightarrow \infty$.
While this is clear for the two larger $U$--values in 
Fig.\ \ref{fig:chi_L}, care must be taken near the transition because
the system appears metallic up to a length scale on the order of
the correlation length which diverges at the transition.
Such a crossover from metallic to insulating behavior is evident in
the $U=4$ curve, for which we have taken 
lattice sizes of up to $L=100$ to show that $\chi$ scales to a
finite value, i.e.\ that the system is insulating.
\begin{figure}
\begin{center}
\epsfig{file=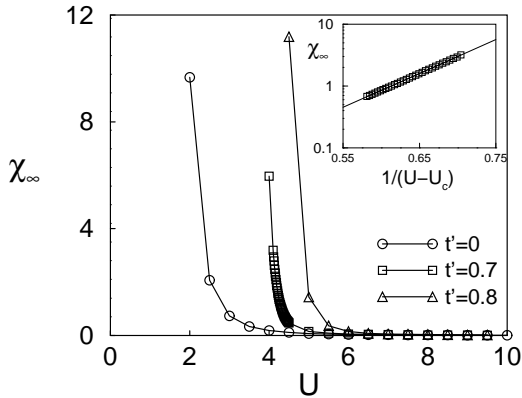,width=7cm}
\end{center}
\caption{Electric susceptibility $\chi_\infty(U,t')$ of the
infinite--size system for $t'=0$,  $0.7$, $0.8$, as a function of $U$;
the lines are guides to the eye. Inset: $\chi_\infty(U,t'=0.7)$ for 
$U=4.1$ to $4.4$ (squares) as a function of $1/(U-U_c)$, on a semilog
scale. 
The line is a fit to an exponential form.  
}
\label{fig:chi_inf}
\end{figure}
In the insulating regime, we expect $\chi$ to be analytic in $1/L$.
We therefore perform finite--size scaling
for large $L$ using a linear fit and extrapolating to $1/L=0$.
The result, $\chi_\infty$, is shown in Fig.\ \ref{fig:chi_inf} for
$t'=0$, 0.7, and 0.8 as a function of $U$.
For $t'=0$, the transition takes place at $U_c=0$, as
discussed previously.
Although we could not obtain a reliable finite--size extrapolation
for $U \lesssim 2$ because the correlation length becomes much
larger than the system sizes we were able to reach, we could observe
numerically that $\chi \sim \Delta^{-2}$ (for $U \lesssim 10$), 
where  $\Delta$ is the charge gap given in 
Ref.\ \onlinecite{ovchin}. 
The extrapolation to $\Delta\rightarrow0$ confirms that $U_c=0$. 
Alternatively, we can fit to the low--$U$ form for $\Delta^{-2}$,
\begin{equation}
	\chi_\infty(t'=0) = \frac{A}{U-U_c} \exp \left[ 
          \frac{B}{(U - U_c)^\sigma}\right]
	\label{eqn:chiexp}
\end{equation}
with the exactly known values $B=4\pi=12.566\ldots$ and
$\sigma=1$;
here the prefactor $1/(U-U_c)$ comes from the logarithmic correction.
This yields $U_c\simeq0.058$ and
we effectively find that $U_c=0$ to within error bars. 
A fit to the form without the logarithmic correction would yield 
$U_c\simeq0.209$,
which is also consistent with zero, but to within a larger error bar.

It is clear from Fig.\ \ref{fig:chi_inf} that the bigger $t'$, the
larger the $U$ at which $\chi$ diverges.
However, one must perform careful fitting in order to accurately
determine $U_c$ and the form of the divergence at $t'>0$, as
an analytical result for the charge gap exists only at $t'=0$.
For $t'=0.7$, we have calculated $\chi$ at many $U$--values near
the transition and have fitted to both power law, $\chi \sim
(U-U_c)^\gamma$ and exponential forms (Eq.\ \ref{eqn:chiexp}), but
without the logarithmic correction.
The logarithmic corrections are, in general, non--universal, i.e.\ 
$t'$--dependent. 
Leaving these corrections out, as argued
above, will only make the determination of $U_c$ less precise.
We find that the fit to the power law form yields $U_c \simeq 3.4$,
a point at which careful finite--size scaling of
$\chi$ yields a finite value of $\chi_\infty$.
Therefore, this $U_c$ is clearly too large.
The exponential fit yields
$\sigma\simeq 1.049$, $B\simeq12.45$ and  
$U_c\simeq2.67$, a more reasonable value of $U_c$.
Note that the values for $\sigma$ and $B$ are again very close to
the ones obtained for $t'=0$.
The inset of Fig.\ \ref{fig:chi_inf} shows a semilog plot of
$\chi_\infty$ versus $1/(U-U_c)$ as well as the fit itself, 
illustrating its good quality. 
We therefore find that the exponential form, Eq.\ (\ref{eqn:chiexp}),
expected in an infinite--order transition, characterizes the
transition at {\em all} $t'$, irrespective of whether a spin gap
exists or whether $U_c$ is finite or zero. 

If hyperscaling is valid, there is
only one relevant length scale $\xi_\infty$ (the
correlation length for $L\rightarrow\infty$) in the vicinity of the quantum 
critical point. This length then determines the
finite--size scaling of the singular part of the ground state
energy density \cite{kim:weichman}
\begin{equation}
	E_0^{\rm sing}/L^d=\xi_\infty^{-(d+z)}f(L/\xi_\infty),	
  \label{E0sing:scaling}
\end{equation}
where $z$ is the dynamic critical exponent and $f$ a universal scaling
function. The quantity
$EL$ is an energy and therefore scales like $\xi_\infty^{-z}$. Using
Eq.(\ref{eqn:susceptibility}), one obtains the scaling behavior of
the electric susceptibility 
\begin{equation}
	\chi=L^{2+z-d}C\Phi(L/\xi_\infty),	\label{chi:fsscaling}
\end{equation}
where $C$ is a
non--universal constant that depends on microscopic details
and $\Phi$ is
a universal function \cite{privman:fisher}. The hyperscaling
assumption also implies that $\Phi$ tends to a finite
value as $L/\xi_\infty\rightarrow0$. This is the region
in which the system appears metallic and in which $\chi$ tends
to \hbox{scale} like $L^2$. Thus $z=1$
is the only consistent value in Eq. (\ref{chi:fsscaling}), in
agreement with exact results for $t'=0$ \cite{stafford}. 
In the opposite limit, $L/\xi_\infty\rightarrow\infty$, the system
behaves as an insulator for all sizes and $\chi$ tends to a finite
value $\chi_\infty$. The scaling form (\ref{chi:fsscaling}) with
$z=d=1$ thus implies $\lim_{x\rightarrow\infty}\Phi(x)=A/x^2$ and
\begin{equation}
	\chi_\infty=\;C\;A\;\;\xi_\infty^2,	\label{chi:scaling} 
\end{equation}
where $A$ is a universal constant.

\begin{figure}
	\begin{center}
	\epsfig{file=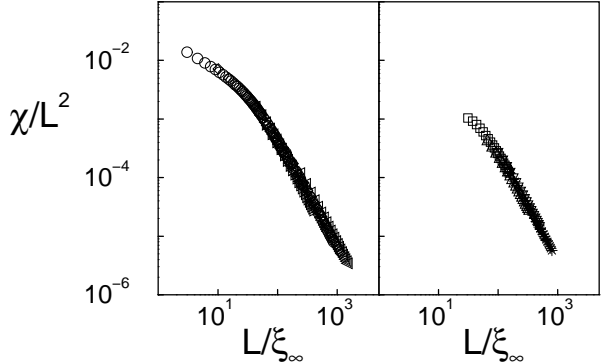,width=8cm}
	\end{center}
	\caption{
	Scaling plots of $\chi(L,U,t')/{L^2}$ versus 
	$L/\xi_\infty(U,t')$ in a log--log scale: $t'=0$ (left),
	$t'=0.8$ (right). Different symbols
	correspond to different values of $U$.
	}
	\label{fig:scaling}
\end{figure}

\begin{figure}
	\begin{center}
		\epsfig{file=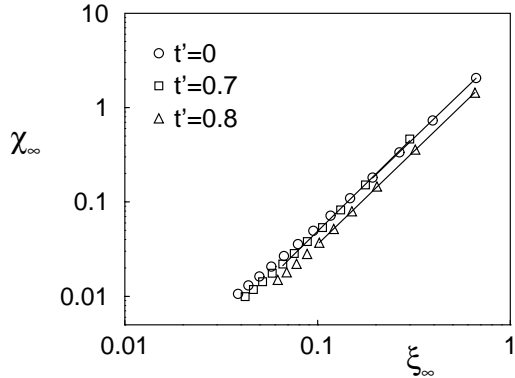,width=6.5cm}
	\end{center}
	\caption{
	Electric susceptibility $\chi_\infty$, versus
	correlation length $\xi_\infty$ for different values of
	$t'$. The lines are power--law fits.
	}
	\label{fig:chi_vs_xi}
\end{figure}

In order to confirm the scaling form Eq.(\ref{chi:fsscaling}) for our
model, we plot the DMRG results
for $\chi/L^2$ as a function of $L/\xi_\infty$ in
Fig.~\ref{fig:scaling}. The quantity $\xi_\infty$ is obtained by
calculating $\xi$ on finite systems using Eq.~(\ref{xi:def}) and then
performing a finite--size extrapolation similar to that used to obtain
$\chi_\infty$. 
Notice that {\em all} $L$ and $U$ points for a
particular $t'$ collapse onto the same curve, confirming
hyperscaling.
Therefore, $\xi_\infty$ behaves as the correlation length, 
which we have checked
by ascertaining that $\xi_\infty$ is the same length (up to a
constant) that characterizes the exponential decay of the
density--density correlation function.
The results of the $1/L$ extrapolation for $\chi$ and $\xi$ are shown
in Fig.~\ref{fig:chi_vs_xi} for three different values of $t'$. A
power--law fit to $\chi_\infty=C'\xi_\infty^{\tilde{\gamma}}$ yields
$\tilde{\gamma}(t'=0)\simeq1.97$, $\tilde{\gamma}(t'=0.7)\simeq2.01$,
$\tilde{\gamma}(t'=0.8)\simeq1.96$, and $C'(t'=0.7)/C'(t'=0)\simeq1$,
$C'(t'=0.8)/C'(t'=0)\simeq0.7$. This confirms the scaling behavior
(\ref{chi:scaling}). It also shows that the constant $C$ depends
weakly on $t'$. 
In summary, our calculations for the
$U-t-t'$ chain at half--filling confirm that the electric
susceptibility $\chi$ (and therefore also the dielectric constant 
$\varepsilon=1+4\pi\chi$)
diverge when approaching the Mott transition from the insulating
side. The polarization fluctuations, which also diverge for
$U\rightarrow U_c$ from above, have been found to be directly
proportional to the correlation length $\xi$ of the Mott insulating
phase. In agreement with the
hyperscaling hypothesis, the metallic or insulating behavior of
the finite--size system depends only on the ratio
$L/\xi_\infty$. The finite--size scaling of $\chi$ can then
be related to a universal scaling function and a dynamic exponent
$z=1$. 
The transition is found to be infinite order and to show the same
critical behavior whether there is a spin gap or not.
As to the origin of this ``dielectric catastrophe'', we conclude, on the
basis of both the inequality $\chi\leq2\xi/\Delta$ and the observed
scaling $\chi_\infty \sim \xi_\infty^2$, that it
involves both a diverging correlation length $\xi$ (linked to the
unbinding of dipoles) and a vanishing of the charge gap $\Delta$.

We thank the Swiss National Foundation for financial support,
under Grant No. 20--53800.98~.
Parts of the numerical computations were done at the Swiss Center for
Scientific Computing.
We thank F.F. Assaad, S. Daul, E. Jeckelmann, G.I. Japaridze,
S. Sachdev, C.A. Stafford and X. Zotos for valuable discussions.

\end{multicols}

\end{document}